\documentclass[epjc3,smallextended,final,twocolumn]{svjour3}

\usepackage{cite} % sort&compress references
\usepackage[breaklinks]{hyperref}%
\usepackage[english]{babel} 
\usepackage[utf8]{inputenc}
\usepackage[OT2,T1]{fontenc}
\usepackage{graphicx}
\usepackage{color}
\usepackage{amsfonts}
\usepackage{bm,bbm}
\usepackage{amsmath,amssymb,amsfonts} 
\usepackage[normalem]{ulem} % to strikeout with \sout

\newcommand{\be}{\begin{equation}}
\newcommand{\ee}{\end{equation}}
\newcommand{\ba}{\begin{eqnarray}}
\newcommand{\ea}{\end{eqnarray}}
\def\bs{\begin{subequations}}
\def\es{\end{subequations}}
\def\a{\alpha}
\def\b{\beta}
\def\de{\delta}
\def\g{\gamma}
\def\la{\lambda}
\def\k{\kappa}
\def\e{\epsilon}

\def\om{\omega}

\def\s{\sigma}

\def\cK{\mathcal{K}}

\def\cP{\mathcal{P}}

\def\ds{d_{\rm S}}

\def\B{\Box}
\newcommand{\Eq}[1]{(\ref{#1})}
\def\com{\color{magenta}}
\def\cob{\color{blue}}

\newcommand{\oarX}[1]{\href{http://arxiv.org/abs/#1}{{\ttfamily\com arXiv:#1}}}
\newcommand{\arX}[1]{\href{http://arxiv.org/abs/#1}{{\ttfamily\com arXiv:#1}}}

\newcommand{\procm}[6]{in \emph{#1}, ed.\ by #2 (#3, #4, #6)}
\newcommand{\doin}[6]{\href{http://dx.doi.org/#1}{\cob  #2 #3 {\bf #4}, #5 (#6)}}
\newcommand{\doinn}[5]{\href{http://dx.doi.org/#1}{\cob  #2 {\bf #3}, #4 (#5)}}
\newcommand{\doij}[5]{\href{http://dx.doi.org/#1}{\cob  #2 {\bf #3}, #4 (#5)}}

\newcommand{\book}[5]{{\it #1} (#2, #3, #5)}

\newcommand{\tia}[1]{#1.}
\newcommand{\boxd}[1]{\boxed{\phantom{\Biggl(}#1\phantom{\Biggl)}}}

\def\rme{\text{e}}
\def\rmd{\text{d}}
\def\rmi{\text{i}}

\journalname{Eur. Phys. J. C}

\date{April 4, 2017}

\begin{document}\sloppy % see http://tex.stackexchange.com/questions/241343/what-is-the-meaning-of-fussy-sloppy-emergencystretch-tolerance-hbadness

\title{Finite entanglement entropy and spectral dimension in quantum gravity}

\author{Michele Arzano\thanksref{addr1,e1} \and Gianluca Calcagni\thanksref{addr2,e2}}
\institute{Dipartimento di Fisica and INFN, ``Sapienza'' University of Rome, P.le A.\ Moro 2, 00185 Rome, Italy\label{addr1}\thankstext{e1}{e-mail: michele.arzano@roma1.infn.it} \and Instituto de Estructura de la Materia, CSIC, Serrano 121, 28006 Madrid, Spain\label{addr2}\thankstext{e2}{e-mail: calcagni@iem.cfmac.csic.es}}

\maketitle

\begin{abstract}
What are the conditions on a field theoretic model leading to a finite entanglement entropy density? We prove two very general results: 1) Ultraviolet finiteness of a theory does not guarantee finiteness of the entropy density; 2) If the spectral dimension of the spatial boundary across which the entropy is calculated is non-negative at all scales, then the entanglement entropy cannot be finite. These conclusions, which we verify in several examples, negatively affect all quantum-gravity models, since their spectral dimension is always positive. Possible ways out are considered, including abandoning the definition of the entanglement entropy in terms of the boundary return probability or admitting an analytic continuation (not a regularization) of the usual definition. In the second case, one can get a finite entanglement entropy density in multi-fractional theories and causal dynamical triangulations.
\end{abstract}

%%%%%%%%%%%%%%%%%%%%%%%%%%%%%%%%%%%%%%%%%%%%%%%%%%%%%%%%%%%%%%%%%%%
%%%%%%%%%%%%%%%%%%%%%%%%%%%%%%%%%%%%%%%%%%%%%%%%%%%%%%%%%%%%%%%%%%%

\section{Introduction: entanglement entropy and return probability}

Quantum correlations in the vacuum state of a quantum field give rise to entanglement entropy once we trace over the degrees of freedom associated, for example, with a region of space. If $\Sigma$ is the boundary of such region (a codimension-2 hypersurface), customary derivations using path-integral techniques in Euclidean time and the replica trick \cite{Casini:2009sr,Nesterov:2010yi,Nesterov:2010jh,Solodukhin:2011gn} lead to an entanglement entropy which is proportional to the ``area'' of the boundary times a divergent integral 
\be\label{edens00}
S \sim A(\Sigma) \int^{+\infty}_{\epsilon^2} \frac{\rmd \s}{\s} \cP_{d} (\s)\,,
\ee
where $\e$ is an ultraviolet (UV) cut-off for small times or coarse resolutions which regularizes the integral and $\s$ is the square of a fictitious diffusion-time parameter, or the inverse of the resolution $\propto 1/\sqrt{\s}$ in a measurement. The function $\cP_{d}(\s)$ in the integrand is the {\it return probability} for a diffusion process on the $(d=D-2)$-dimensional boundary hypersurface, where $D$ is the topological dimension of spacetime. Equivalently, $\cP_{d}(\s)$ is given by the trace of the heat kernel\footnote{In some literature we will invoke later \cite{Akk12}, the name ``heat  kernel'' is reserved to the return probability, but here we follow the convention adopted in the quantum gravity literature.} associated with the propagator of a field theory living on the boundary \cite{Frolov:1998ea}. In particular, \eqref{edens0} represents the effective action of the field restricted to the boundary which, for a free theory, is proportional to the vacuum energy. 

Notice that the integral
\be\label{edens0}
\rho_d^\e=\int^{+\infty}_{\epsilon^2} \frac{\rmd\s}{\s} \cP_{d}(\s)
\ee
can be interpreted as the regularized version of the entanglement entropy {\it density}
\be\label{edens}
\rho_d:=\int^{+\infty}_0\frac{\rmd\s}{\s} \cP_{d}(\s)
\ee
The short-distance correlations responsible for the entanglement entropy are thus intimately related to the properties of a diffusion process on the boundary. Is there a way to formalize this relation? In \cite{Solodukhin:2011gn}, Solodukhin showed that there is indeed a relation between the short-distance scaling of the two-point function and the UV divergence of the entanglement entropy, but their power-law behaviour coincides only for a standard {\it quadratic} energy-momentum dispersion relation.

The relevance of entanglement entropy across a boundary in quantum gravity is well exemplified in Jacobson's derivation of the Einstein equations as an equation of state \cite{Jacobson:1995ab}.  The key assumption of his result is that one can associate a finite (universal) entropy density to any local Rindler horizon and that such entropy is due to the correlations present in the vacuum state of a quantum field. This scenario has led to suggest (see, e.g., \cite{Chirco:2014saa}) that quantum entanglement across the Rindler horizon can be obtained from correlations of the quantum geometry itself and, due to the fundamental discreteness of the geometry, the associated entropy density is finite.

In view of this, it is interesting to investigate whether there exist unconventional quantum field theories which incorporate putative quantum-gravity effects and naturally predict a finite entropy density. Field theories with deformed dispersion relations which either break Lorentz invariance or preserve it (being the deformed dispersion relation a non-linear function of the ordinary Lorentz-invariant mass-shell relation) have already been ruled out in \cite{Nesterov:2010jh} as possible candidates. In that work, the authors showed that such theories can make the short-distance behaviour of the two-point function finite but, no matter how regular this behaviour is, the return probability associated with a deformed dispersion relation {\it always diverges} in the limit $\s\rightarrow 0$, hence the need of a UV cut-off $\e$ in \Eq{edens0}.

In this paper, we turn our attention to more general classes of field-theoretical models and explore whether they lead to a finite entanglement entropy.\footnote{Quantum entanglement is a hot topic on which new papers appear almost on a daily basis. After the submission of this manuscript, we became aware of work on the entanglement entropy with cosmological applications: see \cite{CaLu5} and references therein for more details.} As a starting point, in Sect.\ \ref{sec2} we discuss a theory with compact momentum space, linked to a well-studied model of non-commutative field theory, with all the good UV properties one might desire (a well-behaved Euclidean propagator, calculated here for the first time, and a finite return probability at $\s=0$) which, however, fails to have a finite entanglement entropy density. This counter-example is important because it is a top-down theory that does not fulfill the expectation of a finite entanglement entropy density, despite realizing precisely the features of the \emph{ad hoc} toy model analyzed in \cite{Nesterov:2010jh} and bypassing the no-go result for modified dispersion relations presented there. In Sects.\ \ref{sec3} and \ref{sec4}, we present the main result of this paper, a theorem (proven almost in full) stating a background-independent necessary condition under which a finite entanglement entropy density $\rho_d$ can be obtained:  the spectral dimension $\ds^{\rm b}$ of the spatial boundary of spacetime must be negative. We split the proof in two parts, a technically involved one on the special value $\ds^{\rm b}=0$ (Sect.\ \ref{sec3}) and a much easier one on the case $\ds^{\rm b}>0$ with strict equality (Sect.\ \ref{sec4}).

Retrospectively, from the results of Sect.\ \ref{sec3} it becomes easy to understand the failure of the compact-momentum-space example of Sect.\ \ref{sec2} as due to a violation of this condition. Interestingly, many kinematical states of the class of theories based on a discrete combinatorial structure (group field theory (GFT) and its realizations as loop quantum gravity and spin foams) have a vanishing spectral dimension at the scale of the combinatorial structure \cite{COT3} and, according to our conjecture, their entanglement entropy density is either infinite or zero.

An interesting byproduct of our analysis regards the treatment of the spectral dimension $\ds$ of spacetime. While $\ds$ is traditionally described, in the quantum-gravity literature, in terms of a diffusion process on the spacetime geometry, here we employ the related but essentially alternative view of $\ds$ as the real pole(s) in the spectral zeta function of the theory. Technically, this entails a transformation of the heat kernel from the usual position or momentum space to the Laplace--Mellin momentum space. This tool is very well known in the spectral analysis of fractal geometries \cite{LvF,Tep07,Akk1,Akk12} but it has not received much attention by quantum-gravity researchers. We will show its usefulness in several applications.

This is not the end of the story. An ever more powerful result establishes the impossibility to get a finite entanglement entropy density if the spectral dimension is positive definite (Sect.\ \ref{sec4}). This excludes essentially \emph{all} quantum-gravity and string-related models, where $\ds^{\rm b}\geq 0$, but it also clashes with some positive results obtained with independent techniques, which we will discuss later. To understand such a contradiction, we should look for ways out of this no-go theorem. The most obvious one is to abandon the definition \Eq{edens} and concentrate on other formulations of the entanglement entropy (such as those employed, precisely, in string theory). However, here we propose a more interesting alternative. Without recurring to the regularization \Eq{edens0}, one \emph{can} calculate the integral \Eq{edens} via an analytic continuation in the parameter space of the theory. Once integration is performed, one can revert back to the allowed parameter range and find a finite, positive $\rho_d$. We will show this for the multi-fractional theory with $q$-derivatives ($T_q$ in short) \cite{revmu} as well as for others theories such as causal dynamical triangulations (CDT), and comment on the viability of this solution. 

Our top-down multi-scale examples (non-commutative and multi-fractional) show that, contrary to common wisdom, there is no clear relation between (a) good UV properties and (b) a finite entropy density. The compact-momentum-space case realizes (a) but not (b), while $T_q$ realizes (b) but not (a), at least not in an obvious way. The necessary condition found here does not have the full status of a theorem because its proof entails the properties of the spectral zeta function, which can be studied analytically when the spectral dimension is constant but that become slightly more complicated in a multi-scale geometry (varying spectral dimension, also known as dimensional flow). We will employ an extension to multi-scale geometries that seems very robust. None of this depends on a specific background or dynamics: we use only intrinsic properties of dimensional flow, in the same spirit of recent developments in quantum gravities and multi-fractional theories \cite{revmu,first}.

Before starting, it may be useful to comment on the application of our results to other quantum-gravity models. First, elements such as back-reaction and the renormalization of Newton's constant usually play a role in full calculations of the entanglement entropy at the semi-classical level, while, as we just noted, our approach is more general in the sense that it solely relies on the spectral dimension. The spectral dimension is always well defined in the semi-classical limit, but also in other regimes with heavy non-perturbative quantum effects. This is shown by many examples in the literature (most being fully non-perturbative), among which we mention causal dynamical triangulations, asymptotic safety, loop quantum gravity and spin foams, and string field theory. Therefore, any method employing the spectral dimension may be valid even beyond the semi-classical level, depending on the specific theory. 

Having said that, it is important to specify in which way the method captures quantum-gravity features. We start from the entanglement entropy density \Eq{edens} defined for a generic quantum field theory and we examine under what conditions this quantity is rendered finite. To model quantum-gravity effects, these field theories, representing generic degrees of freedom in two neighboring regions separated by a surface or horizon, have deformed dispersion relations and live on a spacetime with fractal-like features. This abstract framework does not aim to represent full blown quantum gravity with all its details and caveats. However, it does encode dimensional flow, the change of spacetime dimension with the probed scale induced by quantum-gravity effects in all known theories. Dimensional flow is a direct, global manifestation of physics beyond Einstein's gravity and, in particular, is tightly related to microscopic quantum uncertainties in spacetime texture \cite{CaRo2}. Although, as is obvious, dimensional flow does not capture all features of quantum gravity, it is enough to extract interesting phenomenology. In the present case, it is sufficient to draw a consequence regarding the entanglement entropy. In fact, the abstract quantum field theory on which one applies \Eq{edens} can be regarded as an effective description of some degrees of freedom (matter or gravitational) living in a spacetime with varying spectral dimension. The underlying fundamental theory, whatever it is, induces effective dispersion relations or kinetic terms in the effective field theory and, most of all, governs the way the spectral dimension changes. Since no quantum gravity to date predicts a negative spectral dimension, the result of the theorem follows.

By this, we do not mean that the entanglement entropy of the full theory is certainly ill defined. We will consider ways out of this restrictive theorem via possible discreteness or quantum mechanisms in specific theories (Sect.\ \ref{conc}). Our model-independent findings are enough to conclude that dimensional flow \emph{per se} is not sufficient to get a finite results, while we leave the role of discreteness as an open question to be explored in future work.

%%%%%%%%%%%%%%%%%%%%%%%%%%%%%%%%%%%%%%%%%%%%%%%%%%%%%%%%%%%%%%%%%%%
%%%%%%%%%%%%%%%%%%%%%%%%%%%%%%%%%%%%%%%%%%%%%%%%%%%%%%%%%%%%%%%%%%%

\section{Infinite entropy density in a UV-finite example: compact momentum space}\label{sec2}

Before starting our discussion of UV-modified field theories in Euclideanized spacetime, let us recall the relationship between the return probability and the Euclidean two-point function $G(x,y)$ of a scalar theory in ordinary four-dimensional Euclidean space. It is well known that $G(x,y)$ can be written in terms of the heat-kernel $K(x,y;\s)$ as
\be
G(x,y)=-\int_0^{\infty} \rmd \s K(x,y;\s)\,,
\ee
with 
\be
K(x,y;\s)=\frac{1}{(2\pi)^4}\int \rmd^4p\, \rme^{\rmi p_{\mu}(x^{\mu}-y^{\mu})}\, \rme^{-\s C(p)}\,,
\ee
where $C(p)$ is the Fourier transform of the wave operator of the theory which, for a massive scalar field, takes the familiar form $C(p)=p^2+m^2$. The return probability is proportional to trace of the heat kernel:
\be
\cP_{D=4}(\s)\propto K(x,x;\s)=\frac{1}{(2\pi)^4}\int \rmd^4p\, \rme^{-\s C(p)}
\ee
and thus it is related with the short-distance behaviour of the two-point function,
\be
\lim_{x\rightarrow y}G(x,y)\propto -\int_0^{\infty} \rmd \s\, \cP_4(\s)\,.
\ee
Such limit usually diverges as the inverse power of the Euclidean distance between $x$ and $y$. Let us look, for example, at the two-point function of a massive scalar field:
\be
G(x,y)=-\frac{1}{(2\pi)^4}\int \rmd^4p\, \frac{\rme^{\rmi p_{\mu}(x^{\mu}-y^{\mu})}}{p^2+m^2}\,.
\ee
The integral over momentum space can be performed by transforming to spherical coordinates, in particular putting $y=0$. The two-point function can be expressed as 
\be
G(x,0) = \frac{\rmi}{(2\pi)^2} \frac{1}{x} \int^{\infty}_{0} \rmd p\, p^2\, \frac{I_{1}(\rmi px)}{p^2+m^2}\,,
\ee
where $I_{1}(ipx)$ is a modified Bessel function of the first kind, and the final result reads
\be\label{2pmassive}
G(x,0) = -\frac{1}{(2\pi)^2} \frac{m}{x} K_1(m x)\,,
\ee
which, for a massless field, simplifies to 
\be\label{2pmassless}
G(x,0) = -\frac{1}{(2\pi)^2} \frac{1}{x^2}\,.
\ee
Equations \Eq{2pmassive} and \Eq{2pmassless} are divergent in the UV limit $x\rightarrow 0$, as expected. Intuitively, one might think that a theory with a regular two-point function in the UV, reflecting a finite behaviour of quantum correlations at short distances, should also exhibit a finite entanglement entropy density. Probably, the easiest way to modify a field theory in order to get a finite UV behaviour is to introduce ``by hand'' a constant length scale to the distance dependence in the denominator of the two-point function \cite{Padmanabhan:2010wg}. However, it can be shown that such modification amounts to a choice of deformed energy-momentum dispersion relation \cite{Nesterov:2010jh} falling into the class of modified field theories with a diverging return probability and entanglement entropy density.

As a further example of models with an intrinsic UV scale, we hereby consider field theories with a non-trivial geometry of momentum space. This structure naturally introduces a fundamental mass scale in the two-point function without spoiling the Lorentz invariance of the theory. A famous example of such models is a field theory defined on a de Sitter momentum space \cite{AmelinoCamelia:2001fd,Freidel:2007hk,Girelli:2009yz,Arzano:2009ci,Arzano:2010jw,Amelino-Camelia:2013gna}, the momentum-space counterpart of a field theory on $\kappa$-Minkowski non-commutative spacetime, associated with $\kappa$-deformations of relativistic symmetries \cite{AmelinoCamelia:2001fd,Lukierski:1991,Lukierski:1992,Lukierski:1994,Majid:1994,Daszkiewicz:2004xy,KowalskiGlikman:2004tz}. For our purposes, we need to consider the Euclidean version of this model, which we assume to be characterized by a compact momentum space given by a 4-sphere of radius $\kappa$\footnote{See also \cite{Arzano:2016fuy} for another model with compact momentum space not directly related to $\kappa$-deformations.}. The radius $\kappa$ provides a UV energy scale, the Euclidean analogue of the de Sitter momentum-space ``cosmological constant''.\footnote{Notice that an alternative definition of Euclidean $\kappa$-momentum space has appeared in \cite{Arzano:2014jfa}. There, the momentum space is a (non-compact) hyperbolic space on which the Lorentz group acts transitively. In our case, we assume that Lorentz transformation are also ``Euclidean'' and are thus replaced by rotations.} In order to determine whether the geometry of momentum space reflects in a finite UV behaviour of the Euclidean two-point function, we explicitly evaluate the Green function from its integral representation. Focusing for simplicity on the massless case, the two-point function reads
\be
G(x,0) = \frac{\rmi}{(2\pi)^2} \frac{1}{x} \int^{\kappa}_{0} \rmd p\, \, \frac{I_{1}(\rmi px)}{\sqrt{1-\frac{p^2}{\kappa^2}}}\,,
\ee
where the factor $\sqrt{1-{p^2}/{\kappa^2}}$ is obtained from the integration measure over the 4-sphere. Upon evaluation, the integral gives
\be
G(x,0) = -\frac{1}{(2\pi)^2} \frac{1}{x^2} \left[1-\cos (\kappa x)\right]\,,
\ee
which remains finite in the UV limit:
\be
G(0,0)= -\frac{\kappa^2}{2(2\pi)^2}\,, 
\ee
while for large distances $x\rightarrow \infty$ the correlations vanish as expected. This shows that the compact geometry of Euclidean momentum space naturally renders the UV correlations of the theory finite.

One might expect that this finite behaviour in the UV is related to a finite return probability when $\s$ vanishes, which would suggest the possibility of a finite entanglement entropy. In general, in theories with compact momentum space or a non-trivial momentum-space measure the technical reason why this can happen is rather simple. For such models, the return probability is a momentum integral with a certain measure $w(p)$ (non-trivial both in theories with compact momentum space and, e.g., in the multi-scale spacetimes described by multi-fractional theories \cite{revmu}) and a modified dispersion relation $C(p)=p^2\cK(p)$ (where $\cK(p)=1$ in the ordinary case):
%\be\label{repro}
%\cP_D(\s)\propto \int_{-\infty}^{+\infty}\rmd^Dp\,w(p)\,\rme^{-\s p^2\cK(p)}\propto \int_0^{+\infty}\rmd p\,p^{D-1}w(p)\,\rme^{-\s p^2\cK(p)}\,,
%\ee
\ba
\cP_D(\s)&\propto& \int_{-\infty}^{+\infty}\rmd^Dp\,w(p)\,\rme^{-\s p^2\cK(p)}\nonumber\\
&\propto& \int_0^{+\infty}\rmd p\,p^{D-1}w(p)\,\rme^{-\s p^2\cK(p)}\,,\label{repro}
\ea
where the support of $w(p)$ may limit the integration range to a semi-interval or an interval. After the change of variables $k^2:=\s p^2$, one gets
\be
\cP_D(\s)\propto \frac{1}{\s^{D/2}}\int_{-\infty}^{+\infty}\rmd^Dk\,w\left(\frac{k}{\sqrt{\s}}\right)\,\rme^{-k^2\cK(k/\sqrt{\s})}\,.
\ee
Now, in all models related to quantum gravity, $C(\infty)=\infty$ and $C(0)=0$. When $C(p)=p^2$, the integrand is $\s$-independent apart from the measure weight, while in the general case the exponential goes to zero when $\s\to 0$. On the other hand, in all known instances either $w(0)=1$ or, as in compact-momentum-space theories, it has a compact support where the upper limit $\infty$ is replaced by $\sqrt{\s}\kappa$, which collapses to zero when $\s\to 0$. Therefore, in the limit $\s\to 0$ the integral is either $\s$-independent or tends to zero. In the first case, the return probability diverges as $\cP_D(\s)\sim \s^{-D/2}$, while in the second case one has a $0/0$ expression that could give a finite result.

Unfortunately, having a finite $\cP_D(0)$ is not sufficient to have a finite entanglement entropy. In this section, we will show how a known three-dimensional example \cite{Alesci:2011cg} and a four-dimensional model with compact momentum space do have a finite $\cP_D(0)$ but cannot generate a finite entanglement entropy. These counter-examples bar the possibility to cure the divergence in the entanglement entropy by a general mechanism involving, among other possibilities, compact momentum spaces.

In \cite{Alesci:2011cg}, the diffusion process for an $SU(2)$ momentum space, the (Euclideanized) momentum space of a particle coupled to three-dimensional gravity, was studied to determine the associated spectral dimensional flow. In particular, it was found that, in the limit $\s\rightarrow 0$, the associated spectral dimension goes to zero. These results were based on the return probability determined by the heat kernel in $SU(2)$ momentum space,
\be\label{retpro3}
\cP_{D=3}^{\kappa}(\s)=\frac{\sqrt{2}}{4\pi^3}  \int_{|p| \leq \kappa} \frac{\rmd^3\vec{p}}{\sqrt{1-\frac{p^2}{\kappa^2}}} \rme^{-\s \vec{p}^{\, 2}}\,,
\ee
where $\k$ is a momentum UV cut-off related to the radius of the $SU(2)$ group manifold. The integral can be carried out analytically and the result is given in terms of modified Bessel functions of the first kind $I_n(x)$:
\be\label{cp3}
\cP_{D=3}^{\kappa}(\s) =\frac{\sqrt{2}\k^3}{4\pi}\,\rme^{- \frac{\kappa^2\s}{2}} \left[I_0\left(\frac{\kappa^2 \s}{2}\right)-I_1\left(\frac{\kappa^2 \s}{2}\right)\right].
\ee
It is immediate to see that, unlike the return probability in ordinary flat momentum space, $\cP_3^{\kappa}(\s)$ does {\it not} diverge in the limit $\s \rightarrow 0$ and, indeed, we have 
\be\label{cp0}
\cP_{D=3}^{\kappa}(0) = \frac{\sqrt{2}\k^3}{4\pi}\,.
\ee
The normalization in \Eq{retpro3} has been chosen so that the heat kernel is normalized to 1 when integrated over all position space and, consistently, a finite $\k$ corresponds to a finite $\cP_3^{\kappa}(0)$. This should be compared with the standard case in flat space ($\k\to\infty$), where $\cP_{D=3}^\infty(\s)=1/(4\pi\s)^{3/2}$, which diverges when $\s\to 0$.

The same holds for the return probability for a diffusion process on a one-dimensional spacetime boundary, where the momentum space is assumed to be a circular one-dimensional sub-manifold of $SU(2)$: 
\be\label{cp1}
\cP_{d=1}^{\kappa}(\s) = \frac1\pi\int_0^{\kappa} \rmd p\, \rme^{-\s\vec{p}^{\, 2}} = \frac{\text{erf}\left(\kappa \sqrt{\s}\right)}{\sqrt{4\pi\s}}\,,
\ee
for which
\be
\cP_{d=1}^{\kappa}(0) = \frac{\k}{\pi}\,.
\ee
This seems very promising, but the real question is whether this finite behaviour of the return probability suffices to ensure a finite {entropy density}. We can immediately see that this is not the case. Indeed, the integral 
\be
\rho_1=\int_{0}^{\infty} \frac{\rmd \s}{\s}\,\cP_1^{\kappa}(\s)\label{pippo}
\ee
is logarithmically divergent and a UV regulator $\e$ is needed in order to obtain a finite result. 

Analogously, in the example of a momentum space given by a 4-sphere discussed above, we can consider a two-dimensional boundary diffusion process with momentum space given by a 2-sphere:
%\be\label{retpro2}
%\cP_{d=2}^{\kappa}(\s) \propto \int_{|p| \leq \kappa} \frac{\rmd^2\vec{p}}{\sqrt{1-\frac{\vec{p}^2}{\kappa^2}}} \rme^{-\s \vec{p}^{\, 2}}\propto \int^{\kappa}_{0} \frac{p\, \rmd p}{\sqrt{1-\frac{p^2}{\kappa^2}}} \rme^{-\s p^{\, 2}} \propto \frac{F\left(\kappa \sqrt{\s}\right)}{\kappa \sqrt{\s}}\,,
%\ee
\ba
\cP_{d=2}^{\kappa}(\s) &\propto& \int_{|p| \leq \kappa} \frac{\rmd^2\vec{p}}{\sqrt{1-\frac{\vec{p}^2}{\kappa^2}}} \rme^{-\s \vec{p}^{\, 2}}\nonumber\\
&\propto& \int^{\kappa}_{0} \frac{p\, \rmd p}{\sqrt{1-\frac{p^2}{\kappa^2}}} \rme^{-\s p^{\, 2}}\nonumber\\
&\propto& \frac{F\left(\kappa \sqrt{\s}\right)}{\kappa \sqrt{\s}}\,,\label{retpro2}
\ea
where $F$ is the Dawson $F$-function and we left out a $\s$-independent overall factor. This return probability is finite in the short-diffusion-time limit, since
\be
\lim_{\s\rightarrow 0} \frac{F\left(\kappa \sqrt{\s}\right)}{\kappa \sqrt{\s}} = 1\,.
\ee
However, as in the $SU(2)$ momentum-space example above, the entropy density obtained from the Mellin transform of the return probability,
\be\label{entd2}
\rho_2\propto\int^{\infty}_{0}\, \frac{\rmd \s}{\s}\, \frac{F\left(\kappa \sqrt{\s}\right)}{\kappa \sqrt{\s}}\,,
\ee
is divergent and a UV regulator $\e$ is required to render it finite. As in the example of the circle, the divergence goes as $\ln\e$.

To summarize, contrary to naive expectations, having a finite return probability in the limit of diffusion scale going to zero does not guarantee a finite result for the entanglement entropy density. It may be possible that this first approach to the calculation of the entropy density in a deformed field theory is too simple-minded because it relies on two assumptions: (a) the calculation of the entanglement entropy leading to the relation \eqref{edens} between entropy and return probability carries through in a field theory on a curved momentum space in the same way as in an ordinary field theory, and (b) the momentum space for the boundary diffusion process is a compact section of the larger manifold momentum space under consideration. However, both assumptions are quite natural. In the first case, the path integral for a free field theory on the non-trivial momentum spaces considered poses no technical problem \cite{AmelinoCamelia:2001fd} and should go through as in the ordinary case. The main step that, in principle, should require some care would be to make sure that the boundary conditions for the field can be imposed unambiguously. Regarding the second point, we do not really see any particular concern or a plausible alternative approach.

It thus seems that demanding a finite entropy density is quite restrictive and it would be very interesting to explore more in depth what kind of non-standard features a field-theoretical model should exhibit in order to comply with such requirement. For this purpose, we now turn to a different perspective.

%%%%%%%%%%%%%%%%%%%%%%%%%%%%%%%%%%%%%%%%%%%%%%%%%%%%%%%%%%%%%%%%%%%
%%%%%%%%%%%%%%%%%%%%%%%%%%%%%%%%%%%%%%%%%%%%%%%%%%%%%%%%%%%%%%%%%%%

\section{Entanglement entropy and spectral dimension}\label{sec3}

Since the finiteness of the entanglement entropy is related to the analytic properties of the return probability, and since the latter also determines the value of the spectral dimension $\ds$ of the space or spacetime on which the fictitious diffusion process is considered, the study of $\ds$ may offer a new insight in the problem of the entanglement entropy.

%%%%%%%%%%%%%%%%%%%%%%%%%%%%%%%%%%%%%%%%%%%%%%%%%%%%%%%%%%%%%%%%%%%

\subsection{Spectral dimension from the zeta function}

Let us recall that the spectral dimension of spacetime is defined as the scaling of the return probability,
\be\label{ds}
\ds=-2\frac{\rmd\ln\cP_D(\s)}{\rmd\ln\s}\,.
\ee
A constant $\ds$ corresponds to a power law $\cP(\s)\sim \s^{-\ds/2}$; in ordinary Euclidean(ized) space(time), $\ds$ coincides with the topological dimension $D$. We will indicate as $\ds^{\rm b}$ the spectral dimension of the $(D-2)$-dimensional spatial boundary determining the entanglement entropy. Integrating \Eq{ds} and plugging the resulting expression (with $D \to d$) into \Eq{edens}, one gets
\be\label{rod}
\rho_d=\int_0^{+\infty}\frac{\rmd\s}{\s}\exp\left[-\frac12\int^\s\frac{\rmd\s'}{\s'}\,\ds^{\rm b}(\s')\right].
\ee
The intuitive and simple case where $\ds$ is constant illustrates the role of the sign of the spectral dimension. If $\ds^{\rm b}>0$, then $\int^\s \rmd\s'\,\ds^{\rm b}(\s')/\s'=\ds^{\rm b}\ln\s$ and the integral in \Eq{rod} diverges to $-\infty$ at $\s=0$, while if $\ds^{\rm b}<0$ the integral defining $\rho_d$ diverges to $+\infty$ at $\s=+\infty$. %{\sout{However, one cannot conclude that any profile $\ds^{\rm b}(\s)>0$ always requires a regularization at $\s=0$ because $\int^\s \rmd\s'\,\ds^{\rm b}(\s')/\s'$ is an indefinite integral and having a positive integrand does not imply a positive-valued outcome. We will see an example of a non-trivial $\ds^{\rm b}(\s)>0$ with a finite $\rho_d$ in section \ref{sec4}; see eqs.\ \Eq{dsmf} and \Eq{46}.}}

A whole literature has been spent on the dependence of the spectral dimension on the resolution $1/\sqrt{\s}$ as an indicator of the geometric properties of exotic spacetimes. Usually in quantum gravity, Eq.\ \Eq{ds} is the only expression used for the purpose. However, here we take a different and much less explored view, which is more familiar to the fields of fractal geometry and statistical mechanics \cite{Akk12}. Given a function $f(\s)$ living in $\mathbbm{R}$, let us consider its associated \emph{zeta function} $\zeta_f(s):=\int_0^{+\infty}\rmd \s\,\s^{s-1}f(\s)/\Gamma(s)$, where $\Gamma$ is Euler's function. Applying this formula to $\cP_{D}(\s)$, we have the zeta function of the return probability:
\be\label{zetap}
\zeta_D(s):=\zeta_{\cP_{D}}(s)=\frac{1}{\Gamma(s)}\int_0^{+\infty}\rmd \s\,\s^{s-1}\cP_{D}(\s)\,.
\ee
This is just a special way to write the Mellin transform of $\cP_{D}$. Using the identity
\be\nonumber
\frac{1}{\la^{s}}=\frac{1}{\Gamma(s)}\int_0^{+\infty}\rmd \s\,\s^{s-1}\rme^{-\s\la}
\ee
and the definition of the return probability, it is easy to convince oneself that another way to cast the zeta function of the heat kernel is
\be\label{eigev}
\zeta_D(s)=\sum_n\frac{1}{\la_n^s}\,,
\ee
where $\la_n$ are the eigenvalues of the Laplacian. In the continuum case, with the return probability given by \Eq{repro}, and for a non-compact spacetime, $\la_n=p^2\cK(p)$ and
\be\label{zata}
\zeta_D(s)=\Omega_{D}\int_0^{+\infty}\rmd p\,w(p)\,\frac{p^{D-1}}{[p^2\cK(p)]^s}\,,
\ee
where $\Omega_{D}$ is the angular integral.

Now, the zeta function can be analytically continued to the complex plane, where it has poles $s_m\in\mathbbm{C}$. In scale-invariant geometries, the largest real part in this set is nothing but half the spectral dimension $\ds$ in a small-$\s$ expansion of the heat kernel.\footnote{When $\ds$ is constant, it can be defined by $\ds=-2\lim_{\s\to 0}\ln\cP_D(\s)/\ln\s$, which explains the small-$\s$ expansion mentioned in the text. When $\ds$ is not constant, its correct definition is \Eq{ds}.} To understand where this result comes from, we recall that the inverse Mellin transform of the zeta function \Eq{zetap} is
\be\label{cpd}
\cP_D(\s)=\frac{1}{2\pi\rmi}\int_{\e-\rmi\infty}^{\e+\rmi\infty}\rmd s\,\zeta_D(s)\Gamma(s)\,\s^{-s}\,.
\ee
Evaluating it by the residue theorem, we immediately recognize the power-law behavior of the return probability	$\cP_D(\s)\sim\s^{-\ds/2}$ at small $\s$. 

We give two examples of calculation of constant $\ds$ via \Eq{zetap} instead of using \Eq{ds}. In the standard case ($\cK=1=w$), one has $\zeta_D(s)\propto\int_0^{+\infty}\rmd p\,p^{D-1-2s}=I/(D-2s)$, where $I$ is a divergent integral that can be made finite by introducing a regulator $p_\textsc{uv}$. The only pole $s=D/2$ is real, so that $\ds=D$. Another instance is $w=1$ and $\cK=|p|^{2(\g-1)}$, corresponding to a massless dispersion relation $|p|^{2\g}=0$. This type of dispersion relation, where $\g$ is some real parameter, is typical of the UV regime of many quantum gravities and phenomenological models (the full dispersion relation would be of the form $p^2+\b|p|^{2\g}=0$, where $\b$ is a constant). Clearly, the standard case corresponds to $\g=1$. The zeta function of the return probability is
%\be\label{zetaga}
%\zeta_D(s)\propto\int_0^{+\infty}\rmd p\,p^{D-1-2s\g}\to \int_0^{p_\textsc{uv}}\rmd p\,p^{D-1-2s\g} =\frac{p_\textsc{uv}^{D-2s\g}}{D-2s\g}\,,
%\ee
\ba
\zeta_D(s)&\propto&\int_0^{+\infty}\rmd p\,p^{D-1-2s\g}\to \int_0^{p_\textsc{uv}}\rmd p\,p^{D-1-2s\g}\nonumber\\
&=&\frac{p_\textsc{uv}^{D-2s\g}}{D-2s\g}\,,\label{zetaga}
\ea
where $p_\textsc{uv}$ is a UV cut-off introduced to regularize the integral when $D-2s\g>0$. The pole now is at $s={D}/(2\g)={\ds}/{2}$, i.e.,
\be\label{fixedds}
\ds=\frac{D}{\g}\,,
\ee
correctly reproducing the spectral dimension of spacetimes with this modified dispersion relation \cite{SVW2,frc4}. To obtain Eq.\ \Eq{zetaga}, we used Eq.\ \Eq{zata} and integrated in momenta $p$ after performing the integration in $\s$ of \Eq{zetap}. Switching integration order (first on $p$ and then on $\s$), one finds $\cP_D(\s)\propto\s^{-D/(2\g)}$ and $\zeta_D(s)\propto \de[D/(2\g)-s]/\Gamma(s)$, thus getting again Eq.~\Eq{fixedds}. We will come back to this point later.

In no-scale geometries (sets or spaces with constant dimensionality), the poles of the spectral zeta function are known to describe the underlying geometric properties of the set on which $\zeta_D$ is calculated. In particular, the pole with largest real part ${\rm Re}(s)=\ds/2$ determines the spectral dimension $\ds$ in a small-$\s$ expansion \cite{LvF,Tep07,Akk1,Akk12}. In the most general case, $\zeta_D(s)$ has many complex poles and the imaginary part of those such that ${\rm Re}(s_n)=\ds/2$ defines a complex dimension $d_{\rm CS}$. This situation is typical of fractals \cite{Akk12,Akk1,Ast15} and also of multi-fractional spacetimes \cite{revmu,complex}. Here we are going to see that the description of the dimension of spacetime in terms of spectral geometry is valid \emph{only at plateaux in dimensional flow}, i.e., only in those scale ranges where $\ds$ is approximately constant. We argue that in multi-scale geometries (variable $\ds$), one has to consider all the poles of the zeta function, not just the one with largest real value. The examples provided here (the compact-momentum-space model of the previous section and the multi-fractional theory of the next section) strongly indicate a principle which, in the most general terms, can be stated as follows. Meaning by ``poles'' the representative $n,m,\ldots=0$, of the families of poles $s_n$, $s_m$, \dots, we have the
\begin{quote}
{\bf Multi-scale-zeta principle.} \emph{The real parts of the poles of $\zeta_D(s)$ determine the non-vanishing plateaux of the spectral dimension $\ds$ of spacetime.}
\end{quote}
In other words, whenever the spectral dimension of spacetime varies continuously along a resolution (probing) scale $\s$, flat regions in dimensional flow appear ($\ds(\s)$ almost flat and constant in some regions in $\ln\s$). This claim is a conjecture only because we do not have a formal mathematical proof at hand. However, its validity is strongly supported by the observation that the multi-scale-zeta principle is just an application, to all plateau regions in dimensional flow, of the well-known rigorous case where $\ds$ is exactly constant. There are no compelling reasons why the spectral zeta function should not encode all the main values of $\ds$ in a multi-scale geometry. The multi-scale-zeta conjecture agrees with the general notion that the plateaux in dimensional flow carry the main information on the underlying geometry and that transient regions between different plateaux are strongly model-dependent and associated with non-physical features such as regularization schemes \cite{COT3,frc4}.

It is not difficult to recognize an immediate consequence of the multi-scale-zeta principle:
\begin{quote}
{\bf Zero-dimension lemma.} \emph{If $\ds=0$ at some point or plateau in dimensional flow, then $\zeta_D(0)$ is finite.}
\end{quote}
This lemma deals with the only case not explicitly contemplated by the multi-scale-zeta principle. It states that a zero spectral dimension does not correspond to a pole of $\zeta_D(s)$. To prove that $\zeta_D(0)$ remains finite, one would have to show that, in the limit $s\to 0$, the zeta function always has a divergence $\sim\Gamma(s)$ in the numerator, cancelling exactly the factor $\Gamma(s)$ in the denominator in \Eq{zetap}. Spectral theory supports this result. A well-known characteristic of the zeta function is that its pole structure is usually given by a combination of the Riemann zeta (in turn related to Euler gamma function) and of gamma functions. Therefore, it is no wonder that a pole in the numerator would scale as $\Gamma(s)$. The zero-dimension lemma is nothing but a consequence of this cancellation due to the definition \Eq{zetap}.

A multi-scale example satisfying the zero-dimension lemma is the case of the $SU(2)$ momentum space. Taking the Mellin transform of Eq.\ \Eq{cp3}, we get
\be
\zeta_{D=3}(s)=\zeta_{\cP_3^\k}(s)=\frac{\k^{3-2s}\Gamma\left(\tfrac32-s\right)}{\sqrt{2\pi}\pi\Gamma(2-s)}\,,
\ee
which, again, is finite at $s=0$. In this case, the poles are $s_n=3/2+n$, where $n\in\mathbbm{N}$, and the smallest\footnote{Here no small-$\s$ expansion is made.} pole gives the spectral dimension of space $\ds=3$; this is correct only in the IR. The full profile $\ds(\s)$ runs from 0 in the UV up to some peak $>3$ at a mesoscopic scale, then dropping down to 3 in the IR \cite{Alesci:2011cg}.

Another check of the principle and of the lemma will be given in Sect.\ \ref{sec4}.

%%%%%%%%%%%%%%%%%%%%%%%%%%%%%%%%%%%%%%%%%%%%%%%%%%%%%%%%%%%%%%%%%%%

\subsection{Boundary zeta function and entanglement entropy}

Let us go back to the entropy density problem. On the spatial boundary of spacetime, all the above formul\ae\ hold with $D$ replaced by $d=D-2$. The spectral zeta function on the boundary reads
\be\label{zetapb}
\boxd{\zeta_d(s):=\frac{1}{\Gamma(s)}\int_0^{+\infty}\rmd \s\,\s^{s-1}\cP_{d}(\s)\,.}
\ee
Up to a divergent normalization factor $1/\Gamma(0)$, we see that $\zeta_d(0)$ coincides with the entanglement entropy density \Eq{edens}:
\be\label{eefin}
\boxd{\rho_d=\lim_{s\to 0} \Gamma(s)\,\zeta_d(s)\,.}
\ee
The constants $\la_n$ in \Eq{eigev} are the eigenvalues of the Laplacian restricted to the spatial boundary and the real part of the poles of $\zeta_d$ is the spectral dimension $\ds^{\rm b}$ at different plateaux, where ``b'' stands for boundary. For non-compact spaces, the boundary can be regarded as an abstract $(D-2)$-surface. %\footnote{Recall the typical example of a 2-sphere, locally isomorphic to $\mathbbm{R}^2$.} 

It becomes clear now why the examples with compact momentum space did not give rise to a finite $\rho_d$. The one-dimensional momentum space $S^1$ has a zeta function
\be\label{zeta1}
\zeta_1(s)=\zeta_{\cP_1^\k}(s)=\frac{\k^{1-2s}}{\pi(1-2s)}\,.
\ee
Crucially, the factor $1/\Gamma(s)$ in the definition of the zeta function compensates a factor $\Gamma(s)$ coming from integration of the return probability. This is the reason why Eq.\ \Eq{zeta1} is finite at $s=0$ and Eq.\ \Eq{pippo}, which lacks the $1/\Gamma(0)$ prefactor, is not. Incidentally, $\ds^{\rm b}=1$ for this boundary space, but one realizes that this is true only in the IR. The full profile $\ds^{\rm b}(\s)$ obtained from the restriction of \Eq{ds} to the boundary of spacetime runs monotonically from 0 in the UV to 1 in the IR.

Turning our attention to the two-dimensional boundary and transforming Eq.\ \Eq{retpro2}, one has
\be\label{zetadue}
\zeta_2(s)=\zeta_{\cP_2^\k}(s)\propto\frac{\k^{3-2s}\Gamma(1-s)}{\Gamma\left(\tfrac32-s\right)}\,,
\ee
up to an $s$-independent finite normalization constant. This expression diverges at the family of poles $s_n=1+n$, and $n=0$ (the representative of the family) corresponds to $\ds^{\rm b}=2$. Again, $\zeta_2(0)$ is finite while Eq.\ \Eq{entd2} is ill defined.

From this exercise, we learn two things: (a) the entanglement entropy density $\rho_d$ of our compact-space examples always diverges as $\Gamma(s)$ in the limit of Laplace--Mellin momentum $s\to 0$; (b) the zeta function $\zeta_d$ of the heat kernel on the boundary does not have any pole at $s=0$ and thus misses the UV behavior of the spatial geometry thereon ($\ds^{\rm b}=0$ in the UV in these examples). The crucial factor here is the presence of a vanishing spectral dimension, i.e., the existence of a pole in $s=0$ of the zeta function. %The only difference between $\rho_d$ and $\zeta_d$ is a factor $\Gamma(s)$ in the numerator, in the limit $s\to 0$:
 The combination $\Gamma(s)\zeta_d(s)$ in \Eq{eefin} diverges both at the poles of $\zeta_d(s)$ and at $s=0$, provided the latter is not a zero of the zeta function. However if $\zeta_d(0)\propto 1/\Gamma(0)=0$, then Eq.\ \Eq{eefin} is \emph{finite}. Conversely, if $\zeta_d(0)$ is finite or diverges, then the entropy density diverges.

What is the physical meaning of this? When applied to $D\to d$ and Eq.\ \Eq{eefin}, the results obtained so far (multi-scale-zeta principle and zero-dimension lemma) yield a novel and important consequence, a
\begin{quote}
{\bf Necessary condition for a finite entropy density.} \emph{If the entanglement entropy density \Eq{eefin} is finite, then the spectral dimension $\ds^{\rm b}$ of the spatial boundary never vanishes at any scale.}
\end{quote}
This is the physical result we were looking for. It is impossible to have a finite entanglement entropy density when the spectral dimension of the space boundary is zero at some scale.

At this point, we can understand the failure of compact momentum space models in getting a finite entanglement entropy: it is because their spectral dimension vanishes in the UV. Technically, this happens because the zeta function does not scale as $\sim 1/\Gamma(s)$ for small $s$. The fact that the meromorphic function $\zeta_d$ is analytic at $s=0$ leads to a divergent $\rho_d$.

To summarize how we reached the formulation of the above necessary condition, we traded the usual definition \Eq{edens} of the entanglement entropy density for the combination of \Eq{zetapb} and \Eq{eefin}. This is completely equivalent to \Eq{edens} or \Eq{rod}, the only difference being that in \Eq{edens} or \Eq{rod} the limit $s\to 0$ is taken before integrating in $\s$, while in Eq.\ \Eq{eefin} it is taken afterwards. This operation commutes in all the cases considered here (compare Eqs.\ \Eq{pippo}, \Eq{entd2} and \Eq{46} with those that can be found in the limit $\s\to 0$ from, respectively, \Eq{zeta1}, \Eq{zetadue} and \Eq{zetamf}) and it does not entail any apparent problem such as regularization artifacts; in particular, we do not use any regularization such as that in \Eq{edens0}. If $\rho_d$ is finite and non-zero, it will be exactly the same when computed from \Eq{edens}, \Eq{rod}, or \Eq{eefin}. The main result is then obtained from the fact that the spectral dimension is, by definition, the (real part of the) pole of the boundary zeta function $\zeta_d(s)$. In particular, if the boundary spectral dimension vanishes, then $\zeta_d(0)$ diverges. But if $\zeta_d(0)$ diverges, then the right-hand side of Eq.\ \Eq{eefin} cannot be finite.

Let us stress that the non-vanishing $\ds^{\rm b}$ condition is necessary but not sufficient. Even when the spectral dimension does not vanish, it may be possible that the entanglement entropy density is ill defined for other reasons. The case with constant spectral dimension \Eq{fixedds} is a good example of this problem: in \Eq{zetaga} (integration first in $\s$ and then in $p$) we introduced a UV cut-off $p_\textsc{uv}$ to render the zeta function finite and, indeed, the entanglement entropy density is infinite: $\rho_d=\lim_{s\to 0} \Gamma(s)/(d-2s\g)=+\infty$. Integrating first in $p$ and then in $\s$, for $\g$ finite we get zero instead of infinity, $\rho_d=\lim_{s\to 0}\de[D/(2\g)-s]=0$.

%%%%%%%%%%%%%%%%%%%%%%%%%%%%%%%%%%%%%%%%%%%%%%%%%%%%%%%%%%%%%%%%%%%
%%%%%%%%%%%%%%%%%%%%%%%%%%%%%%%%%%%%%%%%%%%%%%%%%%%%%%%%%%%%%%%%%%%

\section{No-go result and ways out}\label{sec4}

After finding, through general arguments and via explicit examples, that UV finiteness is not sufficient and that a non-vanishing spectral dimension is necessary, we discuss a third and more serious obstacle towards a finite entropy density, the following necessary condition: the boundary spectral dimension must be negative definite at short scales. We first prove it and then consider its consequences.

From the definition \Eq{ds},
\be\label{dscont}
\frac{\rmd\cP_D(\s)}{\rmd\s}=-\frac{\ds(\s)}{2\s}\cP_D(\s)\,.
\ee
In essentially all quantum gravity and string related models, $\ds\geq 0$ at all scales. Since $\s\geq 0$ and the return probability is positive definite, then the right-hand side is negative semi-definite, which implies that $\cP_D(\s)$ increases as $\s$ decreases. On the other hand, the integral \Eq{edens} is finite at $\s=0$ provided $\cP_d(\s)$ vanishes faster than $\s$ at short scales. This contradicts the previous condition when applied to the boundary, unless $\ds^{\rm b}(\s)<0$ at $\s\sim 0$. This configuration makes \Eq{edens} finite also at $\s=+\infty$ provided $\cP_d(\s)$ does not grow faster than $\s$ at large scales.

We do not need to review all the calculations of the spectral dimension in quantum gravities to convince the reader that the condition $\ds^{\rm b}(\s)<0$ is unphysical and never realized. However, we cannot accept this result uncritically because we do have at least two examples of theories with finite entanglement entropy density.
\begin{itemize}
\item A class of theories of quantum gravity is based on non-local operators of exponential form $\exp\B$ \cite{Mod1,BGKM,Mod3,CaMo2,MoRa,TaBM}. Kinetic terms in field actions have the typical appearance 
\be\label{nolo}
\sim \phi\B\rme^{-\B/M^2}\phi\,.
\ee
The same type of operators appear also in the gravitational sector and render it non-local. Due to the strong suppression $\sim \rme^{-p^2/M^2}/p^2$ of the free propagator, the spectral dimension of non-local quantum gravity flows to 0 in the UV \cite{CaMo1}, in flagrant violation of the first necessary condition found here. Therefore, we predict a divergent entropy density in models with this type of {\it free} propagator. At the interacting level, the situation is not so clear. The dressed propagator, which is a modification of the free one by interactions, gives rise to a different dispersion relation, hence to a different effective return probability, spectral dimension and entanglement entropy. In \cite{GMRZ}, it is argued that in UV-finite interacting theories characterized by non-local terms of the type \eqref{nolo}, the absence of divergent contributions to the renormalized Newton's constant indeed makes the entropy finite. The key elements here are the interaction terms and UV-finiteness rather than non-locality. %{\cor \sout{It would be very interesting to further explore this scenario to understand which assumption of our no-go argument are evaded in this context.}}
\item The low-energy limit of string field theory has exactly the same type of non-locality \Eq{nolo} but the spectral dimension in the UV is bounded from below by the worldsheet \cite{CaMo1}. Therefore, at short scales
\be
\ds\to 2
\ee
and the spectral dimension never vanishes. From this, according to our spectral analysis the entanglement entropy in string theory may be finite. This statement, as imprecise as it is, agrees with explicit calculations of the entropy regarded as an effect of quantum entanglement \cite{HNTW,HaMa} and with the statistical-mechanics derivation of the black-hole entropy-area law from the counting of microscopic degrees of freedom (BPZ states) \cite{StVa}. Only explicit calculations in certain regimes of the theory, such as those where the AdS/CFT correspondence applies, could verify the above conclusion.
\end{itemize}
Thus, either interactions or different definitions of the entanglement entropy may avoid the no-go theorem. However, instead of giving up the definition \Eq{edens} we propose a third alternative: to continue it analytically whenever possible. This solution is less radical, but perhaps subtler, than inserting an \emph{ad hoc} UV regularization as in expression \Eq{edens0}. For both these reasons, it deserves our attention.

\subsection{Finite entropy density with analytic continuation}

In this section, we illustrate the validity of the multi-scale-zeta principle, of the zero-dimension lemma and of the analytically continued finite-$\rho_d$ necessary condition with a fully analytic example: the multi-fractional theory $T_q$ with $q$-derivatives. This framework, reviewed in \cite{revmu}, is a field theory defined on a multi-fractal geometry with varying Hausdorff and spectral dimension. Dimensional flow follows a multi-parametric universal profile determined only by very general (background- and dynamics-independent) properties of the spacetime dimension \cite{first}. Spacetime is characterized by a hierarchy of length scales $\ell_*\equiv\ell_1>\ell_2>\ldots$; above $\ell_*$, ordinary geometry is recovered, while at scales $\lesssim\ell_*$ the spacetime dimensionality drops below $D$. At ultra-microscopic scales, a discrete structure naturally emerges, encoded in logarithmic oscillations of the geometry. Physical observables are affected by this scale dependence. The generality of the measure, together with the fact that all quantum gravities have dimensional flow, justifies the interest in these theories as an efficient framework wherein to explore all the main physical consequences of dimensional flow and to constrain them with experiments and observations ranging from particle physics to cosmology. There is also the hope to improve the perturbative quantization properties of gravity thanks to dimensional flow itself, as it may happen in other theories.

Depending on the symmetries of the field Lagrangian, there are different versions of multi-fractional dynamics. Here we concentrate on one of the best studied cases, the theory $T_q$. Here the return probability can be calculated exactly and it reads \cite{frc7}
\be\label{gau}
\cP_D(\s)=[4\pi\tau(\s)]^{-\frac{D}{2}}\,,
\ee
where the profile $\tau(\s)$ depends on the scale hierarchy in the same way as the spacetime measure. In general, one can take the polynomial form $\tau(\s)=\sum_{n=1}^N g_n (\s/\s_n)^{\a_n}F_n(\s)$, where $g_n$ are constants, $0<\a_n\leq 1$ are called fractional exponents and $F_n$ is a modulation factor affecting all scales and related to the discrete structure in the UV. Without any loss of information, one can consider a binomial profile with $N=2$ and only one fundamental scale $\s_*=\s_1\propto\ell_*^2$ and one modulation factor $F_\om$ dependent on a frequency parameter $\om$. Overall,
\be\label{taus}
\tau(\s)=\s+\s_*\left(\frac{\s}{\s_*}\right)^\a F_\om(\s),
\ee
where $0<\a<1$. In this paper, we coarse grain over log oscillations, which is equivalent to set $F_\om=1$. Calculating \Eq{ds} with \Eq{gau} and \Eq{taus}, one gets \cite{frc7}
\be\label{dsmf}
\ds(\s)=D\frac{\a+(\s/\s_*)^{1-\a}}{1+(\s/\s_*)^{1-\a}}\,,
\ee
which is positive definite because $\a\geq 0$. 

The calculation of the entanglement entropy density can be done via three completely equivalent routes. We develop all the details of the first to highlight a delicate
analytic continuation one should be careful about;\footnote{This caveat was overlooked in arXiv versions 1 and 2 of the present paper.} the other two routes pass through the same steps. Plugging Eqs.\ \Eq{gau} and \Eq{taus} into \Eq{edens},
\ba
\rho_d &=& \int^{+\infty}_0\frac{\rmd\s}{\s} \frac{1}{\{4\pi[\s+\s_*({\s}/{\s_*})^\a]\}^{\frac{d}{2}}}\nonumber\\
			 %&\stackrel{y:=\s/\s_*}{=}& \frac{1}{(4\pi\s_*)^{\frac{d}{2}}}\int^{+\infty}_0\rmd y \frac{y^{-1}}{(y+y^\a)^{\frac{d}{2}}}\nonumber\\
			 &\stackrel{y:=\s/\s_*}{=}& \frac{1}{(4\pi\s_*)^{\frac{d}{2}}}\int^{+\infty}_0\rmd y \frac{y^{-\frac{d\a}{2}-1}}{(1+y^{1-\a})^{\frac{d}{2}}}\nonumber\\
			 &\stackrel{x:=y^{1-\a}}{=}& \frac{1}{|1-\a|(4\pi\s_*)^{\frac{d}{2}}}\int^{+\infty}_0\rmd x \frac{x^{-\frac{d\a}{2(1-\a)}-1}}{(1+x)^{\frac{d}{2}}},
\ea
where the absolute value accounts for a change of sign of the integral depending on whether $\a\gtrless 1$. This integral coincides with the one in formula 3.194.4 of \cite{GR}, 
%\be\label{inte}
%\int^{+\infty}_0\rmd x \frac{x^{\mu-1}}{(1+\b x)^{\nu}}=\frac{1}{\b^\mu}B(\mu,\nu-\mu)=\frac{1}{\b^\mu}\frac{\Gamma(\mu)\,\Gamma(\nu-\mu)}{\Gamma(\nu)}\,,
%\ee
\ba
\int^{+\infty}_0\rmd x \frac{x^{\mu-1}}{(1+\b x)^{\nu}}&=&\frac{1}{\b^\mu}B(\mu,\nu-\mu)\nonumber\\
&=&\frac{1}{\b^\mu}\frac{\Gamma(\mu)\,\Gamma(\nu-\mu)}{\Gamma(\nu)}\,,\label{inte}
\ea
where $\b=1$, $\nu=d/2$, $\mu=-d\a/[2(1-\a)]$ and $B$ is given in formula 8.384.1. For the integral \Eq{inte} to converge to the right-hand side, it must be $|{\rm arg}\b|<\pi$ and ${\rm Re}\nu>{\rm Re}\mu>0$. The first condition is trivial, while the second requires $1>-\a/(1-\a)>0$, i.e., $\a<0$. Assuming that, one finds
\be\label{46}
\rho_d=\frac{\Gamma\left[\frac{d}{2(1-\a)}\right]\Gamma\left[-\frac{d\a}{2(1-\a)}\right]}{(1-\a)(4\pi\s_*)^{\frac{d}{2}}\Gamma\left(\frac{d}{2}\right)}\,.%\qquad \rho_2=\frac{\Gamma\left(\frac{1}{1-\a}\right)\Gamma\left(-\frac{\a}{1-\a}\right)}{4\pi\s_*(1-\a)}\,,
\ee
Before commenting this expression, let us derive it also from the spectral dimension and from the zeta function. Considering the boundary dimension $\ds^{\rm b}$ and plugging \Eq{dsmf} (with the spacetime dimension $D$ replaced by the boundary dimension $d$; obviously, this is not a regularization) into \Eq{rod}, one obtains \Eq{46}. 
 Setting $F_\om=1$ allows us to find $\zeta_d$ analytically.\footnote{Lifting this assumption makes the pole analysis of the zeta function slightly more complicated and intriguing, although it does not add much to the specific problem of the entanglement entropy. For this reason, we report on that in a separate work \cite{complex}.} 

Alternatively, combining Eqs.\ \Eq{zetap}, \Eq{gau} and \Eq{taus}, and continuing analytically the integral to the regions of the parameter space of the theory, we get
%\be\label{zetamf}
%\zeta_D(s)=\zeta_{D,\a}(s)=\frac{\s_*^s}{(1-\a)(4\pi\s_*)^{\frac{D}{2}}}\frac{\Gamma\left[\frac{D-2s}{2(1-\a)}\right]\Gamma\left[\frac{2s-D\a}{2(1-\a)}\right]}{\Gamma\left(\frac{D}{2}\right)\Gamma(s)}\,.
%\ee
\ba
\zeta_D(s)&=&\zeta_{D,\a}(s)\nonumber\\
&=&\frac{\s_*^s}{(1-\a)(4\pi\s_*)^{\frac{D}{2}}}\frac{\Gamma\left[\frac{D-2s}{2(1-\a)}\right]\Gamma\left[\frac{2s-D\a}{2(1-\a)}\right]}{\Gamma\left(\frac{D}{2}\right)\Gamma(s)}\,.\label{zetamf}
\ea
From \Eq{eefin}, one gets again \Eq{46}. In none of these routes leading to \Eq{46} did we use any regularization, but we did make an analytic continuation to the region $\a<0$.

\begin{figure}
\centering
\includegraphics[width=8.2cm]{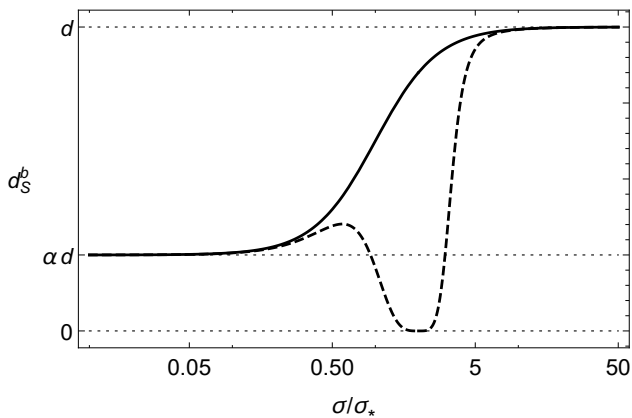}
\caption{\label{fig1} The typical kink profile of the boundary spectral dimension in many quantum gravities (solid curve). This is reproduced by the multi-fractional theory with $q$-derivatives, whose zeta function \Eq{zetamf} with $D\to d$ has poles at $s=d/2=\ds^{\rm b,IR}/2$ and $s=d\a/2=\ds^{\rm b,UV}/2$, where $0<\a<1$. The dashed curve represents a generic \emph{ad hoc} profile of the spectral dimension vanishing at some intermediate scale, leading to an infinite entanglement entropy density.}
\end{figure}

Negative values of $\a$ correspond to a negative spectral dimension, which is consistent with the no-go theorem. At this point, however, we make an experiment: we analytically continue \Eq{46} and \Eq{zetamf} to values $0<\a<1$ and see whether we find a sensible result. There are several consequences we can draw from \Eq{zetamf}.
\begin{itemize}
\item Let $\a\neq 0$. The poles of $\zeta_D$ are $s_n=D/2+n(1-\a)$ and $s_m=D\a/2-m(1-\a)$, where $n,m\in\mathbbm{N}$. For $n=0=m$, we obtain the IR and UV spectral dimension, respectively:
\be\label{iruv}
\ds^{\rm IR}=D\,,\qquad \ds^{\rm UV}=D\a\,,
\ee
in agreement with \Eq{dsmf}. Thus, we have shown that the spectral zeta function brings about information as regards the spectral dimension of both the IR and the UV plateau, when neither of them is zero. Generalizing to many fractional coefficients $\a_n$, one will find as many poles as the number of non-zero plateaux $\ds\simeq D\a_n$. Although we were unable to find an analytic expression of $\zeta_D$ beyond the binomial case, there is really no conceptual obstacle against this expectation. The multi-scale-zeta principle is thus confirmed.
\item $\zeta_D(s)$ diverges in the limit $\a\to 1$, in agreement with the regularized divergence in \Eq{zetaga} with $\g=1$.
\item When $\a=0$, the UV spectral dimension obtained from Eqs.\ \Eq{gau} and \Eq{taus} is $\ds^{\rm UV}=0$, in agreement with Eq.\ \Eq{iruv}. However, this result can no longer be obtained from the zeta function \Eq{zetamf}, which reduces to
\be
\zeta_{D,0}(s)=\frac{\s_*^s}{(4\pi\s_*)^{\frac{D}{2}}}\frac{\Gamma\left(\frac{D}{2}-s\right)}{\Gamma\left(\frac{D}{2}\right)}\,.
\ee
Here we can appreciate the cancellation between the $\Gamma(s)$ factor in the denominator and the factor $\Gamma[(s-D\a/2)/(1-\a)]\to\Gamma(s)$ in the numerator. At $s=0$, $\zeta_{D,0}(0)=(4\pi\s_*)^{-D/2}$ is finite, in agreement with the zero-dimension lemma. Consistently, in the limit $\s_*\to 0$ we get the power-law divergence of \Eq{zetaga} with $\g=1$.
\item For $D\to d$ (zeta function of the boundary diffusion process), the entanglement entropy density is exactly \Eq{46}. For a generic $\a$, this expression is finite, consistently with the fact that the residue of \Eq{cpd} at $s=0$ is zero. The $d$-dimensional entropy density diverges at $\a=1+d/(2n)$ and at $\a=[1+d/(2m)]^{-1}$, where $n,m\in\mathbbm{N}^+$. The first set of poles is excluded because $\a<1$. The second set
\be
\a=\frac{2}{D},\frac{4}{2+D},\frac{6}{4+D},\cdots
\ee
corresponds to geometries with infinite entropy density. Therefore, in $D=4$ the exponents $\a=1/2,2/3,3/4,\ldots$ are ruled out if we insist to have a finite $\rho_2$. This leads to two major results.
\begin{itemize}
\item The value $\a=1/2$, suggested by quantum-gravity arguments and realizing the famous two-dimensional UV limit $\ds^{\rm UV}=2$ so popular in quantum gravity, does not give a finite entanglement entropy. In a sense to be better specified below, this may indicate that it is difficult to get a finite $\rho_d$ in quantum gravity. The compact-momentum-space model and some results mentioned in the introduction \cite{Nesterov:2010jh} support this view.
\item The excluded values are a countable subset of $1/2\leq\a<1$. This range is special in another multi-fractional theory called $T_{\g=\a}$, characterized by multi-scale fractional derivatives in the action and of which $T_q$ is an approximation \cite{revmu}. The spectral dimension in $T_{\g=\a}$ is basically the same as in $T_q$ and we may transpose all the present discussion to that case. In $T_{\g=\a}$, values in the interval $0<\a<1/2$ are excluded if we want to have a normed space at all scales. The limit case $\a=1/2$ is acceptable but it corresponds to a taxicab geometry where the shortest path between two points is not unique \cite{frc1}. On the other hand, when log oscillations are turned on and the full discrete structure of spacetime is considered, the norm condition breaks down in the UV for any $\a$ and one may simply give it up altogether. In fact, space is normed in the IR regardless of what happens in the UV. In turn, log oscillations generate a rather peculiar behaviour of the UV propagator that can have important consequences for the renormalization of gravity not explorable with naive power-counting arguments \cite{revmu}. The bottom line of all this is that, at least in multi-fractional theories but probably also in quantum gravity at large, a finite entanglement entropy density (range $0<\a<1/2$) may be tightly related to geometries highly non-trivial in the UV and where gravity is well behaved.
\end{itemize}
\end{itemize}
Applied to the spatial boundary ($D\to d$), Eq.\ \Eq{iruv} and the general profile (typical of many theories of quantum gravity) derived from Eq.\ \Eq{zetamf} is depicted in Fig.\ \ref{fig1}.

\begin{figure}
\centering
\includegraphics[width=8.2cm]{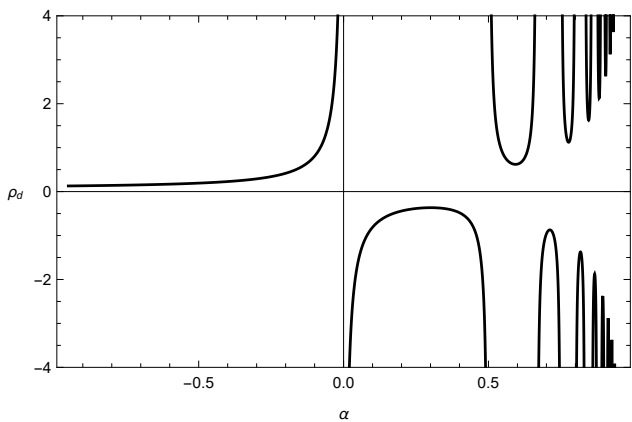}
\caption{\label{fig2} The entanglement entropy density \Eq{46} for $d=2$ (four-dimensional spacetime) and $\s_*=1$.}
\end{figure}

Let us now comment on the analytic continuation $0<\a<1\to\a<0\to 0<\a<1$. Figure \ref{fig2} shows the behaviour of the entanglement entropy density \Eq{46} as a function of $\a$.
As expected, $\rho_d$ is positive definite in the healthy but unphysical case $\a<0$. However, $\rho_d$ is finite and positive also in the intervals
%\be\label{rangea}
%\left(1+\frac{d}{2m}\right)^{-1}<\a<\left(1+\frac{d}{2m+2}\right)^{-1},\qquad m=1,3,5,\dots\,.
%\ee
\ba
&&\left(1+\frac{d}{2m}\right)^{-1}<\a<\left(1+\frac{d}{2m+2}\right)^{-1},\nonumber\\
&& m=1,3,5,\dots\,.\label{rangea}
\ea
Although this might be just an artifact of the analytic continuation, it is intriguing to speculate that such analytic continuation might be simply part of the definition of the entanglement entropy density when calculated via the boundary return probability. If this were true, it would be a genuine intrinsic property of the theory, not a UV regularization as \Eq{edens0}. To check the absence of inconsistencies, one should calculate $\rho_d$ from other definitions, for instance using microstates or thermodynamics, and verify whether the entanglement entropy is really finite in the ranges \Eq{rangea} or if, on the contrary, it diverges for the allowed parameter space. As this would go beyond the scope of this work, we will stop here, but not before noting that this solution would rescue not only the multi-fractional theory with $q$-derivatives, but also other quantum gravity models with a dimensional flow similar to \Eq{dsmf}. In fact, the profile of the spectral dimension of multi-fractional theories with binomial measure, where $\ds(\s)$ runs between two non-zero values, is typical also in many other models of quantum gravity. Quantum gravities such as CDT \cite{AJL4,SVW1,GWZ2,CoJu} and asymptotic safety \cite{LaR5,RSnax} respect our weak necessary condition (their scale-dependent spectral dimension never vanishes), but not the no-go theorem because $\ds>0$ in these cases. However, for the case of CDT we can say something more. The CDT profile of the spectral dimension has a form very similar to \Eq{dsmf}, $\ds^{\rm CDT}(\s)\simeq (b+4\s)/(c+\s)>0$ \cite{AJL4,GWZ2,CoJu}, where $b,c>0$ are constants found numerically. Calculating the entropy density for this profile on a $d$-dimensional surface, $\ds^{\rm b}(\s)\simeq (b+d\s)/(c+\s)$, we have $\rho_d(b,c)\propto\Gamma[-b/(2c)]/\Gamma[d/2-b/(2c)]$. This expression is finite only if $b/(2c)\neq 1$, which is avoided if the theoretical values of $b$ and $c$ are irrational. Again, one goes through an analytic continuation of the parameters.

%%%%%%%%%%%%%%%%%%%%%%%%%%%%%%%%%%%%%%%%%%%%%%%%%%%%%%%%%%%%%%%%%%%
%%%%%%%%%%%%%%%%%%%%%%%%%%%%%%%%%%%%%%%%%%%%%%%%%%%%%%%%%%%%%%%%%%%

\section{Conclusions: role of dimensional flow and discreteness}\label{conc}

In this paper, we have determined a necessary condition to obtain a finite entanglement entropy density $\rho_d$: the spectral dimension of the spatial boundary of the spacetime geometry must not be positive at any scale. If it does, then $\rho_d$ diverges. We have seen the negative example of theories with compact momentum space (necessary condition on non-vanishing spectral dimension violated)\footnote{We should mention here that, in the example \cite{Arzano:2014jfa} of non-compact Euclidean momentum spaces, the spectral dimension never vanishes. However, the entropy density diverges due to the non-compactness of momentum space as in the examples of deformed dispersion relations studied in the literature \cite{Nesterov:2010yi}.} and the example of multi-fractional spacetimes, successful only after augmenting the definition of $\rho_d$ with an analytic continuation. We also considered the case \Eq{fixedds} of continuous scale-invariant geometries (constant spectral dimension), where $\rho_d$ diverges. 

When $\ds^{\rm b}<0$, $\cP_d(\s)\sim \s^{-\ds^{\rm b}/2}$ grows with $\s$ and the spectral dimension makes little sense both as the scaling of the return probability of a diffusing process (which should decrease with the diffusion time) and as an indicator of spacetime dimensionality (which should be positive semi-definite). A physical interpretation of why a vanishing spectral dimension is problematic could be that, when $\ds\to 0$, the boundary effectively collapses to a point. Trying to match the quantum entanglement and statistical-mechanics interpretations of the entropy density, one might identify this type of degenerate configuration as the responsible for an ill-defined counting of degrees of freedom on the boundary, eventually resulting in an infinite entropy density. However, the boundary configuration may be much more complicated than a degenerate point. If quantum geometry hinders diffusion of a particle too much, for instance when its structure at some given scale is discrete or too coarse, then the spectral dimension of spacetime vanishes at that scale and the entropy density blows up. The condition $\ds^{\rm b}= 0$ implies a constant return probability $\cP_d(\s)\sim \s^{-\ds^{\rm b}/2}= {\rm const}$ at the diffusion scales of the zero plateau, meaning that a test particle does not diffuse on the boundary at those scales. This may happen if the boundary geometry is too irregular or disconnected. Therefore, a finite entropy can arise only in geometries sufficiently smooth, non-degenerate, or not pathologically rough. Note that one cannot define the concept of ``irregularity'' as a synonym of ``nowhere differentiable,'' since nowhere-differentiable geometries such as those in the multi-fractional theory $T_\g$ with fractional derivatives correspond to multi-fractal dimensional profiles where $\ds\neq 0$ and, as the above calculation for its approximation $T_q$ shows, the entanglement entropy should be finite. 

We can start to draw some conclusions from the above results. We begin to recognize \emph{dimensional flow} as playing an important role in the mechanism responsible for a finite entropy density. However, as the compact momentum space examples indicates, dimensional flow \emph{per se} is not sufficient to get $\rho_d<\infty$. These geometries are multi-scale but $\rho_d$ is infinite. The same conclusion holds for non-local quantum gravity, as we have seen in the previous section, and unless one considers interactions. On the other hand, string theory avoids the vanishing-$\ds$ problem, while multi-fractional theories and CDT work only if we allow for an analytic continuation. Therefore, there exist theories (all multi-scale, i.e., with varying $\ds$) based on either a nowhere-differentiable or an ordinary differentiable structure that may realize a finite entanglement entropy density. Differentiability is not the main requirement here, the only common element being that the spacetime described by these theories is multi-scale. All these examples are on a continuum and one may wonder whether \emph{discreteness} renders the entropy density finite.
\begin{itemize}
\item A graphical instance of the virtues of discreteness is provided by deterministic fractals. These sets are characterized by a discrete scale invariance and have a constant spectral dimension. Their spectral zeta function, however, does not vanish at $s=0$ \cite{Akk1,Akk12} and, thus, their entanglement entropy density has a divergent term \cite{Ast15}, as in the case \Eq{fixedds} of continuous scale-invariant geometries. Thus, theories of quantum gravity based on a multi-fractal geometry and solving the entropy-density problem might work but not because of an irregular spacetime texture.
\item Multi-fractional spacetimes have a measure with precisely such properties: both $T_q$ and $T_\g$ are field theories on a multi-fractal spacetime with a UV discrete scale invariance. However, as we commented in the previous section, discreteness is not necessary to get $\rho_d<\infty$, since a finite $\rho_d$ can be obtained in the continuum regime of the theory $T_q$ (coarse-grained log oscillations). At the same time, a finite $\rho_d$ is also curiously related to a region of the parameter space of the multi-fractional theory $T_\g$ corresponding to geometries where a continuum description is flawed ($0<\a<1/2$), but this does not strictly imply that discreteness must enter the game.
\item Discreteness plays an important role also in quantum gravities based upon pre-geometric combinatorial structures, such as loop quantum gravity, spin foams and their mother theory GFT. These theories exhibit a finite number of degrees of freedom on boundaries, which reproduce the entropy-area law for black holes \cite{ABCK,ABK,Mei04,OrPS}. This statistical-mechanics interpretation of the entropy density (in terms of state counting) should agree, at some point, with the interpretation of the same quantity as due to the quantum entanglement between states inside and outside the boundary or horizon. The results of \cite{Chirco:2014saa} are encouraging in this respect, but we are now in the position to make an observation that jeopardizes this hope. Numerical investigations of the properties of a wide class of kinematical states of quantum geometry appearing in GFT-spin foams-loop quantum gravity found that the UV is dominated by effects of the underlying discrete combinatorial structure, so that $\ds\to 0$ at short scales \cite{COT3}. Therefore, according to our necessary condition, these quantum geometries do not have a finite entanglement entropy density, in contradiction with the suggestions of \cite{Chirco:2014saa} but in syntony with independent arguments against a resolution of the information-loss problem (presumably related to the entanglement-entropy problem) in loop quantum gravity \cite{Boj14}. Before reaching a final verdict, however, we would like to emphasize that: (a) keeping in mind our cautionary remarks made at the end of Sect.\ \ref{sec2}, we should recall that our conclusion is based on the assumption that the relation \eqref{edens} between entanglement entropy and the return probability of a boundary diffusion process remains valid in the particular quantum-geometry setting under consideration; (b) a resolution of this apparent contradiction might lie in the fact that, for example, the states analyzed in \cite{COT3} are kinematical; states in the dynamical Hilbert space, so far left alone due to technical difficulties, might fare better. The limit $\ds\to 0$ is caused by a discreteness artifact of the theory, the combinatorial structure of complexes. In a way not really clear to us at this stage, it may be possible that such an artifact is smoothed out in fully dynamical geometric states. In other words, if we accepted that no physical measurement can take place at the discreteness scale of underlying simplicial complexes, then the regime $\ds^{\rm UV}=0$ would be unreachable for all purposes, thus avoiding our no-go result. In the case of black holes, this condition would amount to forbid objects with a Planckian event horizon. Future studies in loop quantum gravity should attempt to bypass the necessary condition found here and to address this issue.
\item  Field theories with a hard UV cut-off implemented as a finite spatial ``bandwidth'' based on Shannon sampling theory \cite{Kempf:2009us} share various features with discrete lattice models, including a minimal wavelength for the field oscillators. Recent studies show that also in these theories the presence of a UV cut-off does not lead to a finite entanglement entropy even though the field degrees of freedom in such model seem to occupy an incompressible volume \cite{Pye:2015tta}. From our perspective, it would be of interest to study the information density carried by the field degrees of freedom in the various models analyzed in this work along the lines of \cite{Pye:2015tta}, to see which features are responsible for a finite entanglement entropy density.
\end{itemize}

To conclude, the main agent responsible for a finite entropy density in quantum gravity may be dimensional flow. This factor is not sufficient by itself, as shown by compact-momentum-space and non-local examples. The role of discreteness is less clear, but it might turn out to be a liability rather than an asset. String theory and the multi-fractional example in the absence of log oscillations show that discreteness is not necessary, while conflicting evidence in loop quantum gravity and GFT is questioning discreteness as an efficient agent. To better understand the relationships in this love triangle among entanglement entropy, varying geometry and discrete or irregular structures, we will have to wait for further advances. The most crucial element of disturbance, the positive sign of the spectral dimension forbidding a finite $\rho_d$, kills virtually all quantum-gravity and string-related models. Changing the definition of the entanglement entropy, as we did minimalistically by an analytic continuation, might be advisable but it should be better understood. The use of the spectral zeta function promoted in this work is a promising starting point from where to refocus the problem.

%%%%%%%%%%%%%%%%%%%%%%%%%%%%%%%%%%%%%%%%%%%%%%%%%%%%%%%%%%%%%%%%%%%%%%%
%%%%%%%%%%%%%%%%%%%%%%%%%%%%%%%%%%%%%%%%%%%%%%%%%%%%%%%%%%%%%%%%%%%%%%%

\section*{Acknowledgments} The work of G.C.\ is under a Ram\'on y Cajal contract and is supported by the I+D Grant FIS2014-54800-C2-2-P.

%%%%%%%%%%%%%%%%%%%%%%%%%%%%%%%%%%%%%%%%%%%%%%%%%%%%%%%%%%%%%%%%%%%%%%%
%%%%%%%%%%%%%%%%%%%%%%%%%%%%%%%%%%%%%%%%%%%%%%%%%%%%%%%%%%%%%%%%%%%%%%%


\begin{thebibliography}{99}
%\cite{Casini:2009sr}
\bibitem{Casini:2009sr} H.\ Casini and M.\ Huerta, \tia{Entanglement entropy in free quantum field theory} \doin{10.1088/1751-8113/42/50/504007}{J.\ Phys.}{A}{42}{504007}{2009} [\arX{0905.2562}].
%\cite{Nesterov:2010yi}
\bibitem{Nesterov:2010yi} D.\ Nesterov and S.N.\ Solodukhin, \tia{Gravitational effective action and entanglement entropy in UV modified theories with and without Lorentz symmetry} \doin{10.1016/j.nuclphysb.2010.08.006}{Nucl.\ Phys.}{B}{842}{141}{2011} [\arX{1007.1246}].
%\cite{Nesterov:2010jh}
\bibitem{Nesterov:2010jh} D.\ Nesterov and S.N.\ Solodukhin, \tia{Short-distance regularity of Green's function and UV divergences in entanglement entropy} \doij{10.1007/JHEP09(2010)041}{JHEP}{09}{041}{2010} [\arX{1008.0777}].
%\cite{Solodukhin:2011gn}
\bibitem{Solodukhin:2011gn} S.N.\ Solodukhin, \tia{Entanglement entropy of black holes} \doinn{10.12942/lrr-2011-8}{Living Rev.\ Relativ.}{14}{8}{2011} [\arX{1104.3712}].
\bibitem{Akk12} E.\ Akkermans, \tia{\href{http://www.ams.org/books/conm/601/11962/conm601-11962.pdf}{\cob Statistical mechanics and quantum fields on fractals}} \procm{Fractal Geometry and Dynamical Systems in Pure and Applied Mathematics II: Fractals in Applied Mathematics}{D.\ Carfi, M.L.\ Lapidus, E.P.J.\ Pearse and M.\ van Frankenhuijsen}{AMS}{Providence}{U.S.A.}{2013} [\arX{1210.6763}].
%\cite{Frolov:1998ea}
\bibitem{Frolov:1998ea} V.P.\ Frolov and D.\ Fursaev, \tia{Black hole entropy in induced gravity: reduction to 2D quantum field theory on the horizon} \doin{10.1103/PhysRevD.58.124009}{Phys.\ Rev.}{D}{58}{124009}{1998} [\oarX{hep-th/9806078}].
%\bibitem{BH1} J.D.\ Bekenstein, \tia{Black holes and entropy} \doin{10.1103/PhysRevD.7.2333}{Phys.\ Rev.}{D}{7}{2333}{1973}.
%\bibitem{BH2} S.W.\ Hawking, \tia{Particle creation by black holes} \doinn{10.1007/BF02345020}{Commun.\ Math.\ Phys.}{43}{199}{1975}; \doinn{10.1007/BF01608497}{}{46}{206}{1976}.
%\cite{Kovtun:2003wp}
%\bibitem{Kovtun:2003wp} P.\ Kovtun, D.T.\ Son and A.O.\ Starinets, \tia{Holography and hydrodynamics: diffusion on stretched horizons} \doij{10.1088/1126-6708/2003/10/064}{JHEP}{0310}{064}{2003} [\oarX{hep-th/0309213}].
%\cite{Chirco:2010xx}
%\bibitem{Chirco:2010xx} G.\ Chirco, C.\ Eling and S.\ Liberati, \tia{Universal viscosity to entropy density ratio from entanglement} \doin{10.1103/PhysRevD.82.024010}{Phys.\ Rev.}{D}{82}{024010}{2010} [\arX{1005.0475}].
%\cite{Jacobson:1995ab}
\bibitem{Jacobson:1995ab} T.\ Jacobson, \tia{Thermodynamics of space-time: the Einstein equation of state} \doinn{10.1103/PhysRevLett.75.1260}{Phys.\ Rev.\ Lett.}{75}{1260}{1995} [\oarX{gr-qc/9504004}].
%\cite{Chirco:2014saa}
\bibitem{Chirco:2014saa} G.\ Chirco, H.M.\ Haggard, A.\ Riello and C.\ Rovelli, \tia{Spacetime thermodynamics without hidden degrees of freedom} \doin{10.1103/PhysRevD.90.044044}{Phys.\ Rev.}{D}{90}{044044}{2014} [\arX{1401.5262}].
\bibitem{CaLu5} S.\ Capozziello and O.\ Luongo, \tia{Information entropy and dark energy evolution} \arX{1704.00195}.
\bibitem{COT3}  G.\ Calcagni, D.\ Oriti and J.\ Th\"urigen, \tia{Dimensional flow in discrete quantum geometries} \doin{10.1103/PhysRevD.91.084047}{Phys.\ Rev.}{D}{91}{084047}{2015} [\arX{1412.8390}].
\bibitem{revmu} G.\ Calcagni, \tia{Multifractional theories: an unconventional review} \doij{10.1007/JHEP03(2017)138}{JHEP}{03}{138}{2017} [\arX{1612.05632}].
\bibitem{first} G.\ Calcagni, \tia{Multiscale spacetimes from first principles} \doin{10.1103/PhysRevD.95.064057}{Phys.\ Rev.}{D}{95}{064057}{2017} [\arX{1609.02776}].
\bibitem{CaRo2} G.\ Calcagni and M.\ Ronco, \doin{10.1016/j.nuclphysb.2017.07.016}{Nucl.\ Phys.}{B}{923}{144}{2017} [\arX{1706.02159}].
\bibitem{LvF}   M.L.\ Lapidus and M.\ van Frankenhuysen, \book{Fractal Geometry, Complex Dimensions and Zeta Functions}{Springer}{New York}{U.S.A.}{2006}.
\bibitem{Tep07} A.\ Teplyaev, \tia{Spectral zeta functions of fractals and the complex dynamics of polynomials} \doinn{10.1090/S0002-9947-07-04150-5}{Trans.\ Am.\ Math.\ Soc.}{359}{4339}{2007} [\oarX{math/0505546}].
\bibitem{Akk1}  E.\ Akkermans, G.V.\ Dunne and A.\ Teplyaev, \tia{Physical consequences of complex dimensions of fractals} \doinn{10.1209/0295-5075/88/40007}{Europhys.\ Lett.}{88}{40007}{2009} [\arX{0903.3681}].
\bibitem{Padmanabhan:2010wg} T.\ Padmanabhan, \tia{Finite entanglement entropy from the zero-point-area of spacetime} \doin{10.1103/PhysRevD.82.124025}{Phys.\ Rev.}{D}{82}{124025}{2010} [\arX{1007.5066}].
\bibitem{Girelli:2009yz} F.\ Girelli, E.R.\ Livine and D.\ Oriti, \tia{Four-dimensional deformed special relativity from group field theories} \doin{10.1103/PhysRevD.81.024015}{Phys.\ Rev.}{D}{81}{024015}{2010} [\arX{0903.3475}].
\bibitem{Arzano:2009ci} M.\ Arzano, J.\ Kowalski-Glikman and A.\ Walkus, \tia{Lorentz invariant field theory on $\kappa$-Minkowski space} \doinn{10.1088/0264-9381/27/2/025012}{Class.\ Quantum Gravity}{27}{025012}{2010} [\arX{0908.1974}].
\bibitem{Arzano:2010jw} M.\ Arzano, \tia{Anatomy of a deformed symmetry: field quantization on curved momentum space} \doin{10.1103/PhysRevD.83.025025}{Phys.\ Rev.}{D}{83}{025025}{2011} [\arX{1009.1097}].
\bibitem{Amelino-Camelia:2013gna} G.\ Amelino-Camelia, M.\ Arzano, G.\ Gubitosi and J.\ Magueijo, \tia{Dimensional reduction in momentum space and scale invariant cosmological fluctuations} \doin{10.1103/PhysRevD.88.103524}{Phys.\ Rev.}{D}{88}{103524}{2013} [\arX{arXiv:1309.3999}].
\bibitem{Freidel:2007hk} L.\ Freidel, J.\ Kowalski-Glikman and S.\ Nowak, \tia{Field theory on kappa-Minkowski space revisited: Noether charges and breaking of Lorentz symmetry} \doin{10.1142/S0217751X08040421}{Int.\ J.\ Mod.\ Phys.}{A}{23}{2687}{2008} [\arX{0706.3658}].
\bibitem{AmelinoCamelia:2001fd} G.\ Amelino-Camelia and M.\ Arzano, \tia{Coproduct and star product in field theories on Lie algebra noncommutative space-times} \doin{10.1103/PhysRevD.65.084044}{Phys.\ Rev.}{D}{65}{084044}{2002} [\oarX{hep-th/0105120}].
\bibitem{Lukierski:1991} J.\ Lukierski, H.\ Ruegg, A.\ Nowicki and V.N.\ Tolstoi, \tia{$q$-deformation of Poincar\'e algebra} \doin{10.1016/0370-2693(91)90358-W}{Phys.\ Lett.}{B}{264}{331}{1991}.  
\bibitem{Lukierski:1992} J.\ Lukierski, A.\ Nowicki and H.\ Ruegg, \tia{New quantum Poincar\'e algebra and $\kappa$-deformed field theory} \doin{10.1016/0370-2693(92)90894-A}{Phys.\ Lett.}{B}{293}{344}{1992}.
\bibitem{Lukierski:1994} J.\ Lukierski, H.\ Ruegg and A.\ Nowicki, \tia{Quantum deformations of nonsemisimple algebras: the example of $D=4$ inhomogeneous rotations} \doinn{10.1063/1.530526}{J.\ Math.\ Phys.}{35}{2607}{1994}.
\bibitem{Majid:1994} S.\ Majid, H.\ Ruegg, \tia{Bicrossproduct structure of $\kappa$-Poincar\'e group and noncommutative geometry} \doin{10.1016/0370-2693(94)90699-8}{Phys.\ Lett.}{B}{334}{348}{1994} [\oarX{hep-th/9405107}].
\bibitem{Daszkiewicz:2004xy} M.\ Daszkiewicz, K.\ Imilkowska, J.\ Kowalski-Glikman and S.\ Nowak, \tia{Scalar field theory on kappa-Minkowski space-time and doubly special relativity} \doin{10.1142/S0217751X0502238X}{Int.\ J.\ Mod.\ Phys.}{A}{20}{4925}{2005} [\oarX{hep-th/0410058}].
\bibitem{KowalskiGlikman:2004tz} J.\ Kowalski-Glikman and S.\ Nowak, \tia{Quantum $\kappa$-Poincar\'e algebra from de Sitter space of momenta} \oarX{hep-th/0411154}.
\bibitem{Arzano:2016fuy} M.\ Arzano and F.\ Nettel, \tia{UV dimensional reduction to two from group valued momenta} \doin{10.1016/j.physletb.2017.02.005}{Phys.\ Lett.}{B} {767}{236}{2017} [\oarX{1611.10343}].
\bibitem{Arzano:2014jfa} M.\ Arzano and T.\ Trze\'sniewski, \tia{Diffusion on $\kappa$-Minkowski space} \doin{10.1103/PhysRevD.89.124024}{Phys.\ Rev.}{D}{89}{124024}{2014} [\arX{1404.4762}].
\bibitem{Alesci:2011cg} E.\ Alesci and M.\ Arzano, \tia{Anomalous dimension in semiclassical gravity} \doin{10.1016/j.physletb.2011.12.026}{Phys.\ Lett.}{B}{707}{272}{2012} [\arX{1108.1507}].  
\bibitem{SVW2}  T.P.\ Sotiriou, M.\ Visser and S.\ Weinfurtner, \tia{From dispersion relations to spectral dimension---and back again} \doin{10.1103/PhysRevD.84.104018}{Phys.\ Rev.}{D}{84}{104018}{2011} [\arX{1105.6098}].
\bibitem{frc4}  G.\ Calcagni, \tia{Diffusion in multiscale spacetimes} \doin{10.1103/PhysRevE.87.012123}{Phys.\ Rev.}{E}{87}{012123}{2013} [\arX{1205.5046}].
\bibitem{Ast15} A.\ Faraji Astaneh, \tia{Entanglement entropy on fractals} \doin{10.1103/PhysRevD.93.066004}{Phys.\ Rev.}{D}{93}{066004}{2016} [\arX{1511.01330}].
\bibitem{complex} G.\ Calcagni, \tia{Complex dimensions and their observability} \arX{1705.01619}.
\bibitem{frc7}  G.\ Calcagni and G.\ Nardelli, \tia{Spectral dimension and diffusion in multiscale spacetimes} \doin{10.1103/PhysRevD.88.124025}{Phys.\ Rev.}{D}{88}{124025}{2013} [\arX{1304.2709}].
\bibitem{frc1}  G.\ Calcagni, \tia{Geometry of fractional spaces} \doinn{10.4310/ATMP.2012.v16.n2.a5}{Adv.\ Theor.\ Math.\ Phys.}{16}{549}{2012} [\arX{1106.5787}].
\bibitem{Mod1}  L.\ Modesto, \tia{Super-renormalizable quantum gravity} \doin{10.1103/PhysRevD.86.044005}{Phys.\ Rev.}{D}{86}{044005}{2012} [\arX{1107.2403}].
\bibitem{BGKM}  T.\ Biswas, E.\ Gerwick, T.\ Koivisto and A.\ Mazumdar, \tia{Towards singularity- and ghost-free theories of gravity} \doinn{10.1103/PhysRevLett.108.031101}{Phys.\ Rev.\ Lett.}{108}{031101}{2012} [\arX{1110.5249}].
\bibitem{Mod3}  L.\ Modesto, \tia{Super-renormalizable multidimensional gravity: theory and applications} \doinn{10.1080/21672857.2013.11519717}{Astron.\ Rev.}{8}{4}{2013} [\arX{1202.3151}].
\bibitem{CaMo2} G.\ Calcagni and L.\ Modesto, \tia{Nonlocal quantum gravity and M-theory} \doin{10.1103/PhysRevD.91.124059}{Phys.\ Rev.}{D}{91}{124059}{2015} [\arX{1404.2137}].
\bibitem{MoRa}  L.\ Modesto and L.\ Rachwa\l, \tia{Super-renormalizable and finite gravitational theories} \doin{10.1016/j.nuclphysb.2014.10.015}{Nucl.\ Phys.}{B}{889}{228}{2014} [\arX{1407.8036}].
\bibitem{TaBM}  S.\ Talaganis, T.\ Biswas and A.\ Mazumdar, \tia{Towards understanding the ultraviolet behavior of quantum loops in infinite-derivative theories of gravity} \doinn{10.1088/0264-9381/32/21/215017}{Class.\ Quantum Gravity}{32}{215017}{2015} [\arX{1412.3467}].
\bibitem{CaMo1} G.\ Calcagni and L.\ Modesto, \tia{Nonlocality in string theory} \doin{10.1088/1751-8113/47/35/355402}{J.\ Phys.}{A}{47}{355402}{2014} [\arX{1310.4957}].
\bibitem{GMRZ}  S.\ Giaccari, L.\ Modesto, L.\ Rachwa\l\ and Y.\ Zhu, \tia{Finite entanglement entropy of black holes} \arX{1512.06206}.
\bibitem{HNTW}  S.\ He, T.\ Numasawa, T.\ Takayanagi and K.\ Watanabe, \tia{Notes on entanglement entropy in string theory} \doij{10.1007/JHEP05(2015)106}{JHEP}{05}{106}{2015} [\arX{1412.5606}].
\bibitem{HaMa}  S.A.\ Hartnoll and E.\ Mazenc, \tia{Entanglement entropy in two-dimensional string theory} \doinn{10.1103/PhysRevLett.115.121602}{Phys.\ Rev.\ Lett.}{115}{121602}{2015} [\arX{1504.07985}].
\bibitem{StVa}  A.\ Strominger and C.\ Vafa, \tia{Microscopic origin of the Bekenstein--Hawking entropy} \doin{10.1016/0370-2693(96)00345-0}{Phys.\ Lett.}{B}{379}{99}{1996} [\oarX{hep-th/9601029}].
\bibitem{AJL4}  J.\ Ambj{\o}rn, J.\ Jurkiewicz and R.\ Loll, \tia{Spectral dimension of the universe} \doinn{10.1103/PhysRevLett.95.171301}{Phys.\ Rev.\ Lett.}{95}{171301}{2005} [\oarX{hep-th/0505113}].
\bibitem{SVW1}  T.P.\ Sotiriou, M.\ Visser and S.\ Weinfurtner, \tia{Spectral dimension as a probe of the ultraviolet continuum regime of causal dynamical triangulations} \doinn{10.1103/PhysRevLett.107.131303}{Phys.\ Rev.\ Lett.}{107}{131303}{2011} [\arX{1105.5646}].
\bibitem{GWZ2}  G.\  Giasemidis, J.F.\ Wheater and S.\ Zohren, \tia{Multigraph models for causal quantum gravity and scale dependent spectral dimension} \doin{10.1088/1751-8113/45/35/355001}{J.\ Phys.}{A}{45}{355001}{2012} [\arX{1202.6322}].
\bibitem{CoJu}  D.N.\ Coumbe and J.\ Jurkiewicz, \tia{Evidence for asymptotic safety from dimensional reduction in causal dynamical triangulations} \doij{10.1007/JHEP03(2015)151}{JHEP}{03}{151}{2015} [\arX{1411.7712}].
\bibitem{LaR5}  O.\ Lauscher and M.\ Reuter, \tia{Fractal spacetime structure in asymptotically safe gravity} \doij{10.1088/1126-6708/2005/10/050}{JHEP}{10}{050}{2005} [\oarX{hep-th/0508202}].
\bibitem{RSnax} M.\ Reuter and F.\ Saueressig, \tia{Asymptotic safety, fractals, and cosmology} \doinn{10.1007/978-3-642-33036-0_8}{Lect.\ Notes Phys.}{863}{185}{2013} [\arX{1205.5431}].
\bibitem{ABCK}  A.\ Ashtekar, J.C.\ Baez, A.\ Corichi and K.\ Krasnov, \tia{Quantum geometry and black hole entropy} \doinn{10.1103/PhysRevLett.80.904}{Phys.\ Rev.\ Lett.}{80}{904}{1998} [\oarX{gr-qc/9710007}].
\bibitem{ABK}   A.\ Ashtekar, J.C.\ Baez and K.\ Krasnov, \tia{Quantum geometry of isolated horizons and black hole entropy} \doinn{10.4310/ATMP.2000.v4.n1.a1}{Adv.\ Theor.\ Math.\ Phys.}{4}{1}{2000} [\oarX{gr-qc/0005126}].
\bibitem{Mei04} K.A.\ Meissner, \tia{Black hole entropy in loop quantum gravity} \doinn{10.1088/0264-9381/21/22/015}{Class.\ Quantum Gravity}{21}{5245}{2004} [\oarX{gr-qc/0407052}].
\bibitem{OrPS}  D.\ Oriti, D.\ Pranzetti and L.\ Sindoni, \tia{Horizon entropy from quantum gravity condensates} \doinn{10.1103/PhysRevLett.116.211301}{Phys.\ Rev.\ Lett.}{116}{211301}{2016} [\arX{1510.06991}].
\bibitem{Boj14} M.\ Bojowald, \tia{Information loss, made worse by quantum gravity?} \doinn{10.3389/fphy.2015.00033}{Front.\ Phys.}{3}{33}{2015} [\arX{1409.3157}].
\bibitem{Kempf:2009us}  A.\ Kempf, \tia{Information-theoretic natural ultraviolet cutoff for spacetime} \doinn{10.1103/PhysRevLett.103.231301}{Phys.\ Rev.\ Lett.}{103}{231301}{2009} [\arX{0908.3061}].
\bibitem{Pye:2015tta}   J.\ Pye, W.\ Donnelly and A.\ Kempf, \tia{Locality and entanglement in bandlimited quantum field theory} \doin{10.1103/PhysRevD.92.105022}{Phys.\ Rev.}{D}{92}{105022}{2015} [\arX{1508.05953}].
\bibitem{GR}    I.S.\ Gradshteyn, I.M.\ Ryzhik, \book{Table of Integrals, Series, and Products}{Academic Press}{London}{U.K.}{2007}.
\end{thebibliography}
\end{document}